\documentclass[10pt,twocolumn,letterpaper]{article}

\usepackage{cvpr}
\usepackage{times}
\usepackage{epsfig}
\usepackage{graphicx}
\usepackage{amsmath}
\usepackage{amssymb}
\usepackage{subfigure}
\usepackage{times,amsmath,epsfig}
\usepackage{graphicx}
\usepackage{epstopdf}
\usepackage{comment}

\usepackage[breaklinks=true,bookmarks=false]{hyperref}

\cvprfinalcopy 

\def\cvprPaperID{****} 

\begin{document}

\title{Deep Image Compression via End-to-End Learning}

\author{Haojie Liu$^\dag$, Tong Chen$^\dag$, Qiu Shen, Tao Yue, and Zhan Ma\\
School of Electronic Science and Engineering, Nanjing University, Jiangsu, China\\
\{haojie, tong\}@smail.nju.edu.cn, \{shenqiu, yuetao, mazhan\}@nju.edu.cn\\
$^\dag${\it Authors contributed equally.}\\
}

\maketitle

\begin{abstract}
  We present a lossy image compression method based on deep convolutional neural networks (CNNs), which outperforms the existing BPG, WebP, JPEG2000 and JPEG as measured via multi-scale structural similarity (MS-SSIM), at the same bit rate. Currently, most of the CNNs based approaches train the network using a $l$-2 loss between the reconstructions and the ground-truths in the pixel domain, which leads to over-smoothing results and visual quality degradation especially at a very low bit rate. Therefore, we improve the subjective quality with the combination of a perception loss and an adversarial loss additionally. To achieve better rate-distortion optimization (RDO),  we also introduce an easy-to-hard transfer learning when adding quantization error and rate constraint.  Finally,  we evaluate our method on public Kodak and the Test Dataset P/M released by the Computer Vision Lab of ETH Zurich, resulting in averaged 7.81\% and 19.1\% BD-rate reduction over BPG, respectively.
\end{abstract}

\section{Introduction}

Images record the visual scene of our natural world and are often
compressed for efficient network exchange and local storage. The most well known image compression algorithms are JPEG and its successors JPEG 2000. In the meantime, there are other alternatives
such as WebP, BPG\footnote{An image compression method uses the modified HEVC Intra Profile.}, and so on, shown quite impressive performance gains to further reduce the image size at the same quality. However, all of them present annoying artifacts (e.g., ringing, blocking, etc) at a high compression ratio, resulting in unpleasant user experience. With the exponential growth of the multimedia data, it is inevitable to develop another lossy image compression algorithms with higher performance (a.k.a., using tiny compressed size but presenting high-quality reconstruction).

Recent works have revealed the great potential in lossy image compression using deep learning~\cite{toderici2017full,balle2016end}. These methods utilize a single autoencoder or recurrent autoencoders to generate
feature maps (fMaps) at the bottleneck layer for subsequent quantization and entropy coding.
Quantization induced error/distortion would be utilized for end-to-end optimization.

For instance, Toderici {\it et al.}~\cite{toderici2017full} have applied the Recurrent Neural Network (RNN) to produce entropy-coded bits progressively and to generate layered image reconstructions at different quality scales. Ball\'e {\it et al.}~\cite{balle2016end} have tried to optimize both distortion loss and entropy (rate) loss to improve the overall compression efficiency. In the meantime, Li {\it et al.}~\cite{li2017learning} have proposed a content-weighted approach to further optimize the compressed bit rates.
All aforementioned methods have demonstrated the outstanding coding efficiency that outperform JPEG and JPEG2000 objectively and subjectively. Note that the objective metric utilized for distortion/quality evaluation is the multi-scale structural similarity (MS-SSIM) because of its superior correlation with the subjective opinion score.

On the other hand, Generative Adversarial Networks (GAN) and perceptual loss based approaches~\cite{johnson2016perceptual} have shown a great success in generating images with better visual quality. Instead of calculating distortions directly in pixel domain, these methods measure the similarity in high-level feature domain using a discriminator network or a pre-trained VGG network to mimic the discriminative characteristics of the human visual system (HVS).

Therefore, in this work, we have proposed a deep CNN, deep residual network specifically~\cite{he2016deep}, based image compression scheme that optimizes the end-to-end rate-distortion performance of image compression jointly. Overall structure is consisted of a forward encoder, a quantizer, a backward decoder,  a rate-distortion optimization (i.e., rate estimation and distortion measurement) and a visual enhancement subsystems.

To ensure the fast convergence of the deep neural network, we train the network progressively via transfer learning, i.e., using the networked trained at light compression (lower quantization) to learn the network at higher compression ratio (higher quantization). In the meantime, motivated by the aforementioned perceptual enhancements using GAN and VGGnet, we have also introduced the  perception loss and adversarial loss into the end-to-end optimization pipeline to generate texture and sharp details for noticeable visual quality improvement.

Compared with BPG (with input source sampled at YUV 4:2:0), our method has presented an impressive performance improvement with averaged 7.81\% BD-Rate reduction (i.e., BD-Rate is measured using the MS-SSIM and Bits Per Pixel) on Kodak dataset, and averaged 19.1\% on the Test Dataset P/M released by the Computer Vision Lab of ETH Zurich.

\section{End-to-End Learning Framework}

\begin{figure}[t]
	\centering
	\includegraphics[scale=0.4]{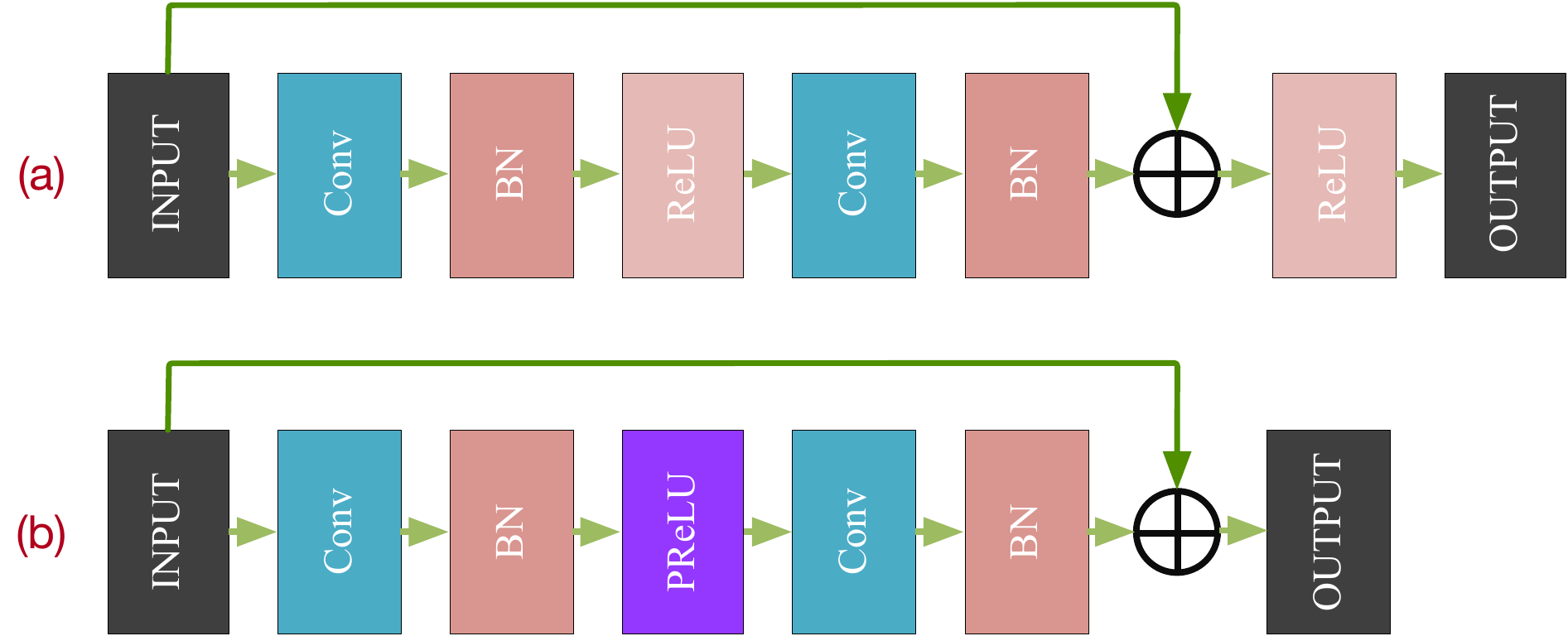}
	\caption{Illustration of a residual unit, (a) the default architecture~\cite{he2016deep}. (b) our proposed architecture with ReLU modifications. }
	\label{fig:resnet}
\end{figure}

We utilize the deep residual network (ResNet)~\cite{he2016deep} in our framework
because of its superior efficiency and fast convergence. To achieve a better convergence
speed, we replace the default rectified linear units (ReLU) with the parametric rectified linear units (PReLU) and remove the nonlinear mapping after the short connection for each residual unit, as shown in Fig.~\ref{fig:resnet}.

\begin{figure*}[t]
	\centering
	\includegraphics[scale=0.34]{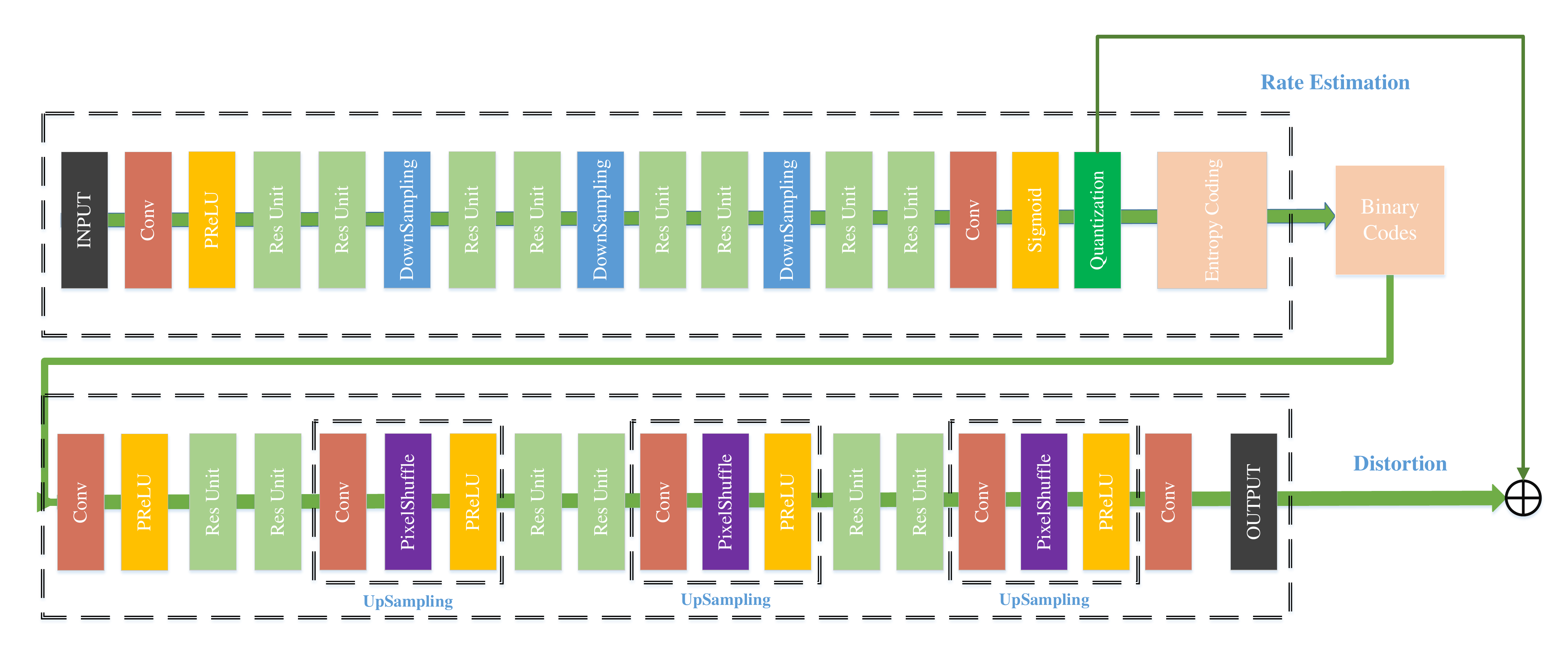}
	\caption{The entire end-to-end learning framework is consisted of a encoder, a quantizer, a decoder and a rate-distortion optimization subsystem with rate estimation and distortion measurement.}
	\label{fig:Network}
\end{figure*}

Overall, the forward encoder network contains eight residual units as shown in Fig.~\ref{fig:Network} and all the down-sampled operations are using a stride-2 4$\times$4 convolutional layer. The decoder has a symmetrical architecture to reconstruct the signal from the compressed fMaps. We choose the pixel-shuffle layers as up-sampled operations considering its decent performances in super resolution (SR). The other convolutional layers all have 3$\times$3 kernel size except the first and last layer using 5$\times$5.

In addition, we use a rate estimation module to approximate the derivable rate loss for back propagation during the training step. The amount of the output features of encoder determine the upper limit of the rate. To adapt images with different content, we dynamically control the bit rates by using different network models.
In one word, we apply less compression on images (i.e., more fMaps) with rich details and vice versa.

\subsection{Quantization and Entropy Coding}
A simple scalar quantization is employed first to reduce the number of bits for representing the extracted fMaps in encoder. First, we scale all the floating coefficients to 6-bit integer via,
\begin{align}
X_Q = {\tt Round}(X_E*({2^Q-1})),
\label{Eq1}
\end{align}
where $X_E$ $\in$ (0, 1) represents the coefficient value of fMaps after the sigmoid activation, $Q$ is the quantization level and set to 6. Then we applied PAQ (a lossless entropy coding method) for the quantized feature cofficients $X_Q$ to generate the binary stream. Then the de-quantized feature coefficients $X_Q/({2^Q-1})$ is fed into the decoder to finally reconstruct the image signals.
We just skip the $\tt Round$ function during backpropagation.

\subsection{Rate Estimation}
The actual bitrates depend on the entropy of the quantized feature maps.
We propose to apply the Lagrangian optimization framework to jointly consider the rate loss $L_R$ and $l$-2 distortion loss.
Here the rate loss $L_R$ is defined as
\begin{align}
L_R = -\mathbb{E} [\log_2 P_q],
\label{Eq2}
\end{align}
where $L_R$ is the entropy approximation of the fMaps at bottleneck layer. Since the derivatives of the quantization function are  almost zero,
we apply a piecewise linear approximation of the discrete $P_q$ to ensure it continuous and differentiable.
Note that Ball\'e \cite{balle2016end} also applied similar idea to do joint rate and distortion optimization.

\subsection{Network Training}
\begin{figure}[b]
	\centering
	\includegraphics[scale=0.4]{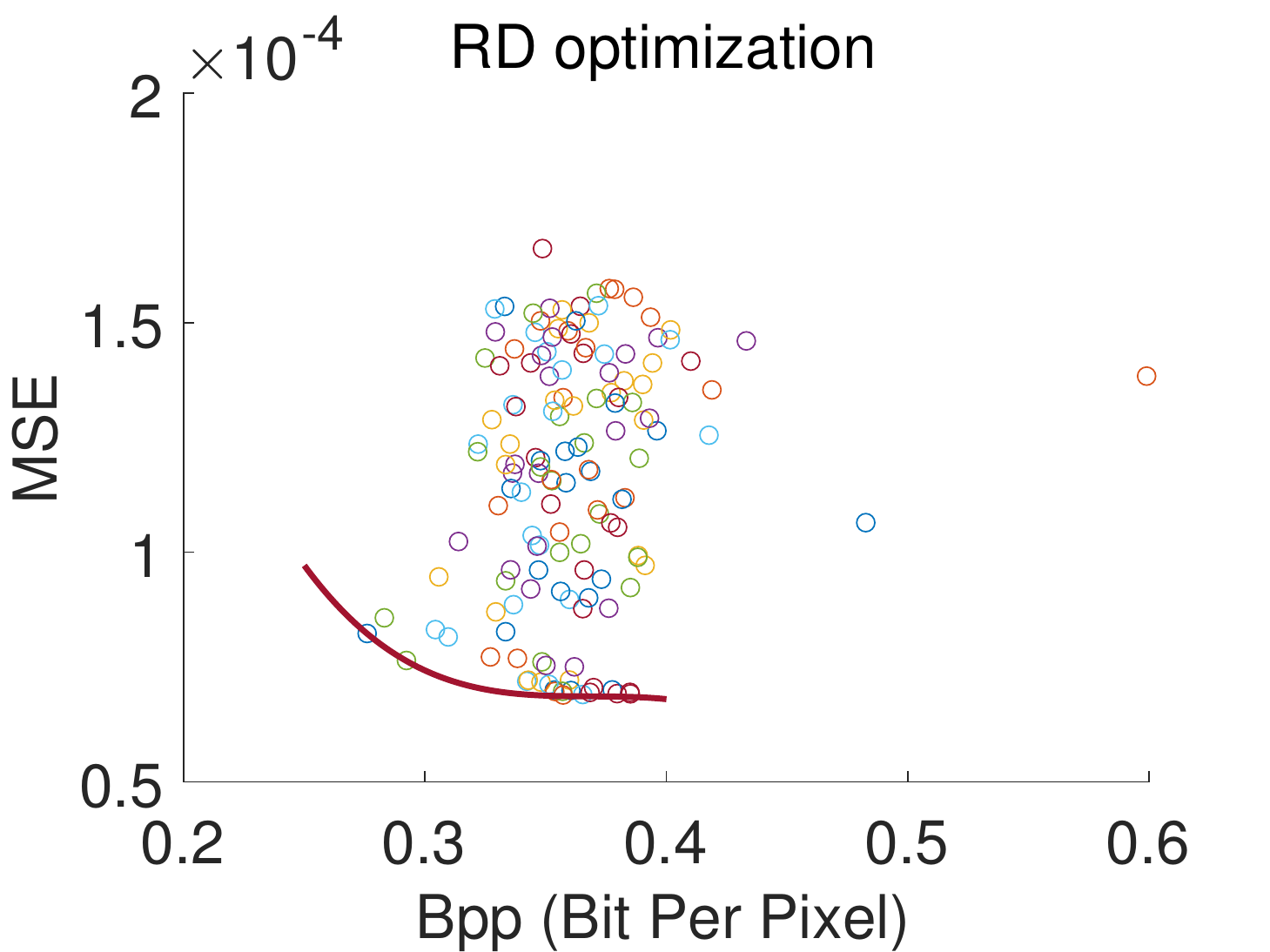}
	\caption{Illustrations of the different rate and distortion generated by different $\lambda$ with a fixed amount of features.  We can do a rate-distortion optimization (RDO) on these discrete points to get the optimized red curve that covers all points on its up-right side.}
	\label{fig:RDO}
\end{figure}

Here, we present more details on how to train CNNs used in the work. In practice, we use the open source data sets released by the Computer Vision Lab of ETH Zurich in CLIC competition. All the images in the training sets are split into 128x128 patches randomly with a data augmentation method such as rotation or scaling, resulting in  80000 patches in total. The objective of training is to minimize the following loss function:
\begin{align}
L = \frac{1}{N}\sum_{n=1}^{N}||Y_n - X_n||^2+ \lambda L_R,
\end{align}
where $X_n$ is the input image, $Y_n$ is the decoded image, $N$ represents the batch size.
We introduce the parameter $\lambda$ to control the penalty of rate loss $L_R$ which is generated from the rate estimation module as shown in Eq.~\eqref{Eq2}.
Inspired by transfer learning, we also apply an easy-to-hard learning method mentioned in the deblocking method named ARCNN and first set $\lambda$ to 0. Without any rate control, we use the optimizer Adam (an adaptive learning rate method) with the learning rate 0.0001 to make fast convergence and generate the pre-training model first after 100 epochs. Then we increase the value of $\lambda$ with a interval of 0.0001 from 0 to 0.002 every 5 epochs to progressively improve the level of rate constrain. Finally, it generates a lot of models with different rate and distortion to make a sophisticated RD optimization as shown in Fig.~\ref{fig:RDO}.

\subsection{Visual Enhancement}
We use another two loss functions to improve the subjective quality of the reconstructed image especially at low bit rate.
\subsubsection{Perception Loss}
Previous works have showed that optimizing the distortion in the feature domain can obviously increase the perceptual information. We choose the last convolutional layer of the 31-layer VGGnet~\cite{simonyan2014very} for feature extraction. The following Eq.~\ref{Eq3} define the perception loss:
\begin{align}
L_{percept} = \frac{1}{N}\sum_{n=1}^{N}||\Psi(Y_n) -\Psi(X_n)||^2,
\label{Eq3}
\end{align}
where $\Psi$ is the VGGnet to compute the features.
\subsubsection{Adversarial Loss}
We introduce another loss following the spirit of GANs. Then, a discriminated neural network $D$ is established to distinguish whether a image is real or fake. For easier training, we replace the DCGAN with the improved Wasserstein GAN (WGAN)~\cite{arjovsky2017wasserstein} to achieve faster convergence and more stable performance. 

The WGAN uses an Earth-Move divergence to measure the similarity of two probabilities and enforce the generator to generate more realistic images. We add the new measurement into the loss function to encourage the reconstructed image having a high probability:
\begin{align}
L_{generator} = - D(Y_n),
\end{align}
The $D$ loss is defined in Eq.~\eqref{Eq6}:
\begin{align}
L_{discriminator} =  D(Y_n)-D(X_n)+\beta L_{penalty},
\label{Eq6}
\end{align}
where $D$ is the discriminative neural network, $L_{penalty}$ is the penalty term mentioned in the improved WGAN, $\beta$ is the parameter of the $L_{penalty}$. Here we set it to 10.

In the end, we merge the different loss functions to build the final measurement component:
\begin{align}
L_{final} = L2 + \lambda_1 L_R+\lambda_2 L_{percept}+\lambda_3 L_{generator},
\end{align}
with $\lambda_2$ = 0.003 and $\lambda_3$ = 0.0001.


\section{Performance Evaluation}
\begin{figure}[t]
	\centering
	\includegraphics[scale=0.34]{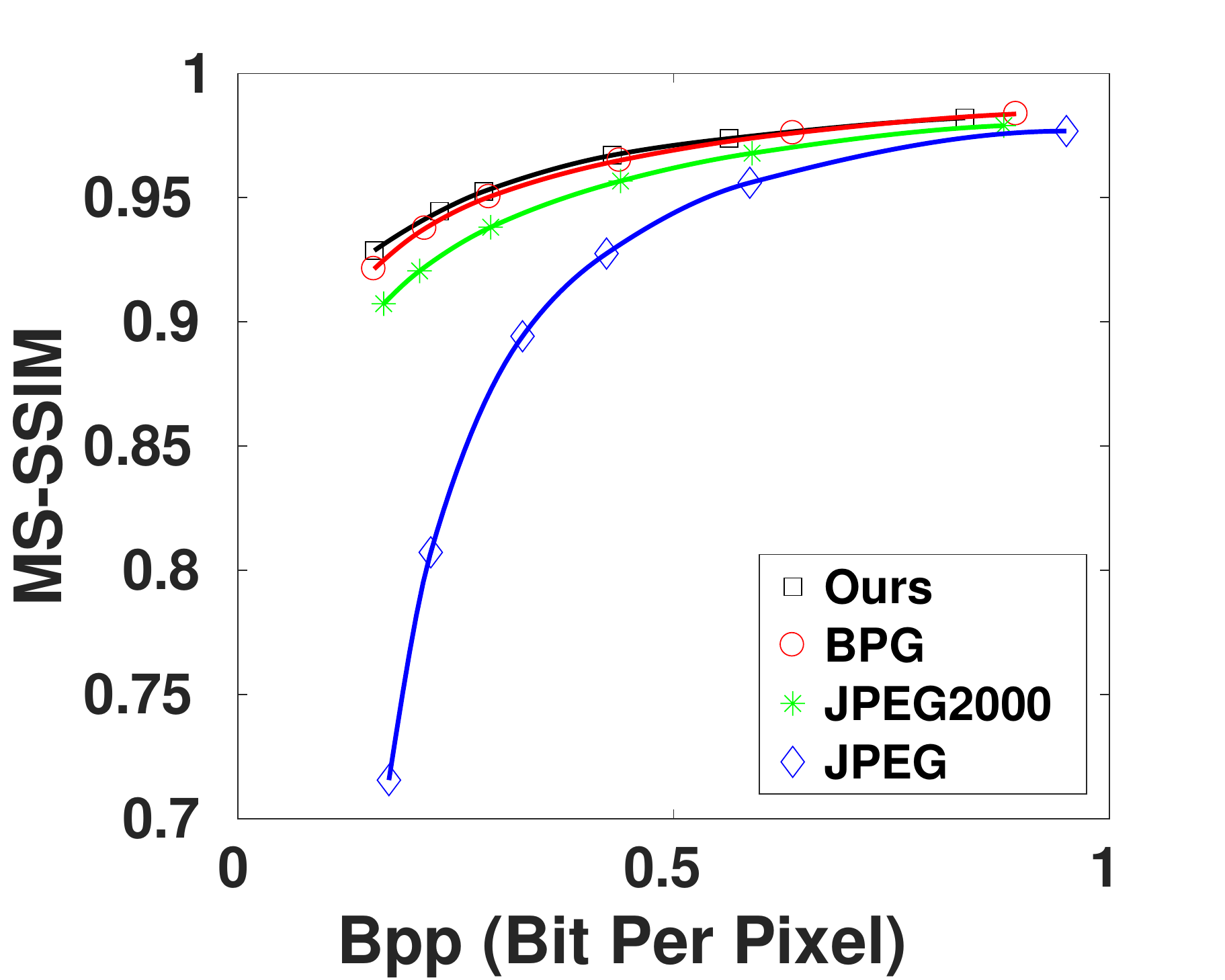}
	\caption{Compression performance on Kodak Dataset, measured in RGB domain, compared with JPEG, JPEG2000 and BPG}
	\label{fig:Kodak}
\end{figure}

\begin{figure}[t]
\centering
\subfigure[casey-fyfe-3340]{\includegraphics[scale=0.053]{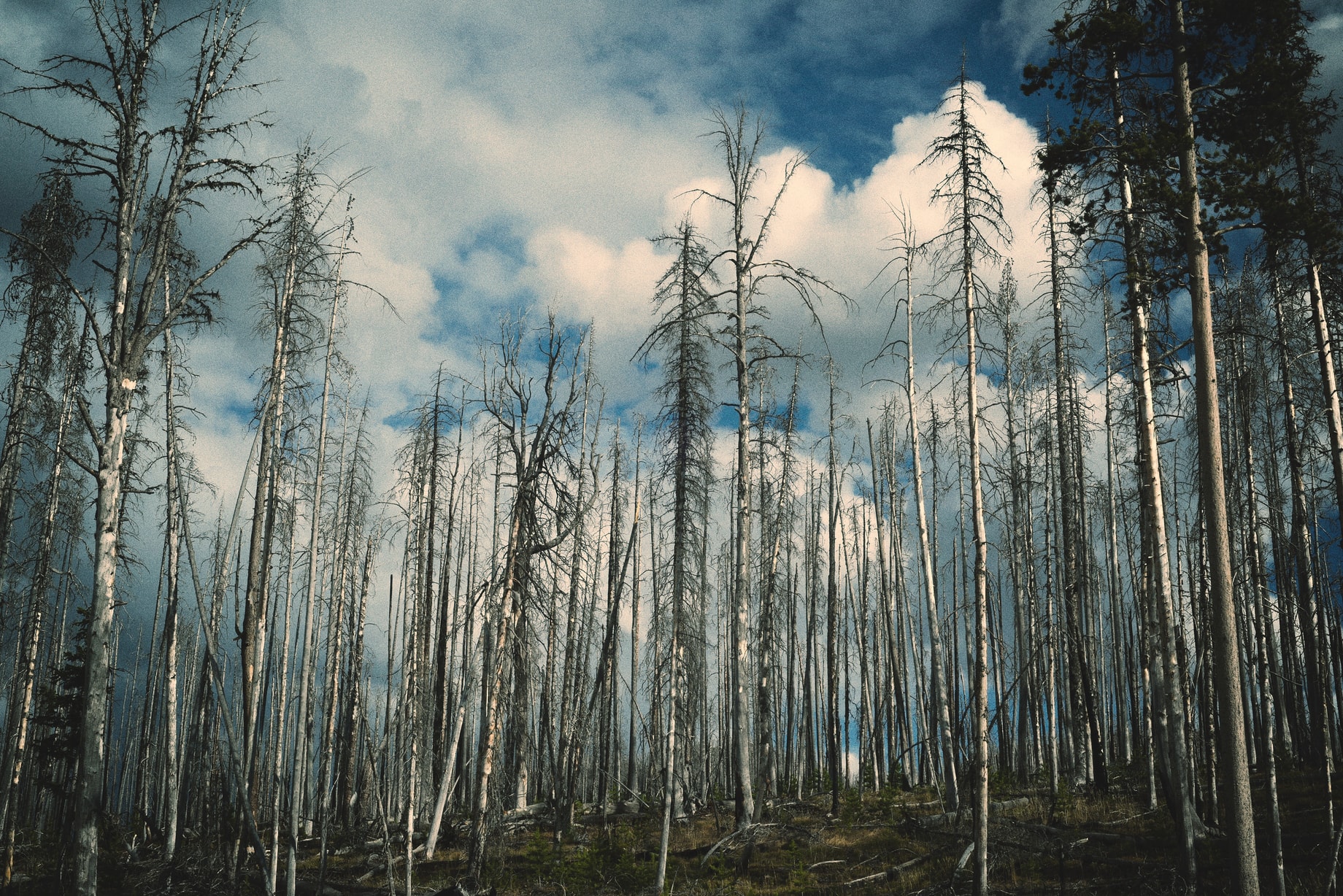}}
\subfigure[lou-levit-369]{\includegraphics[scale=0.0478]{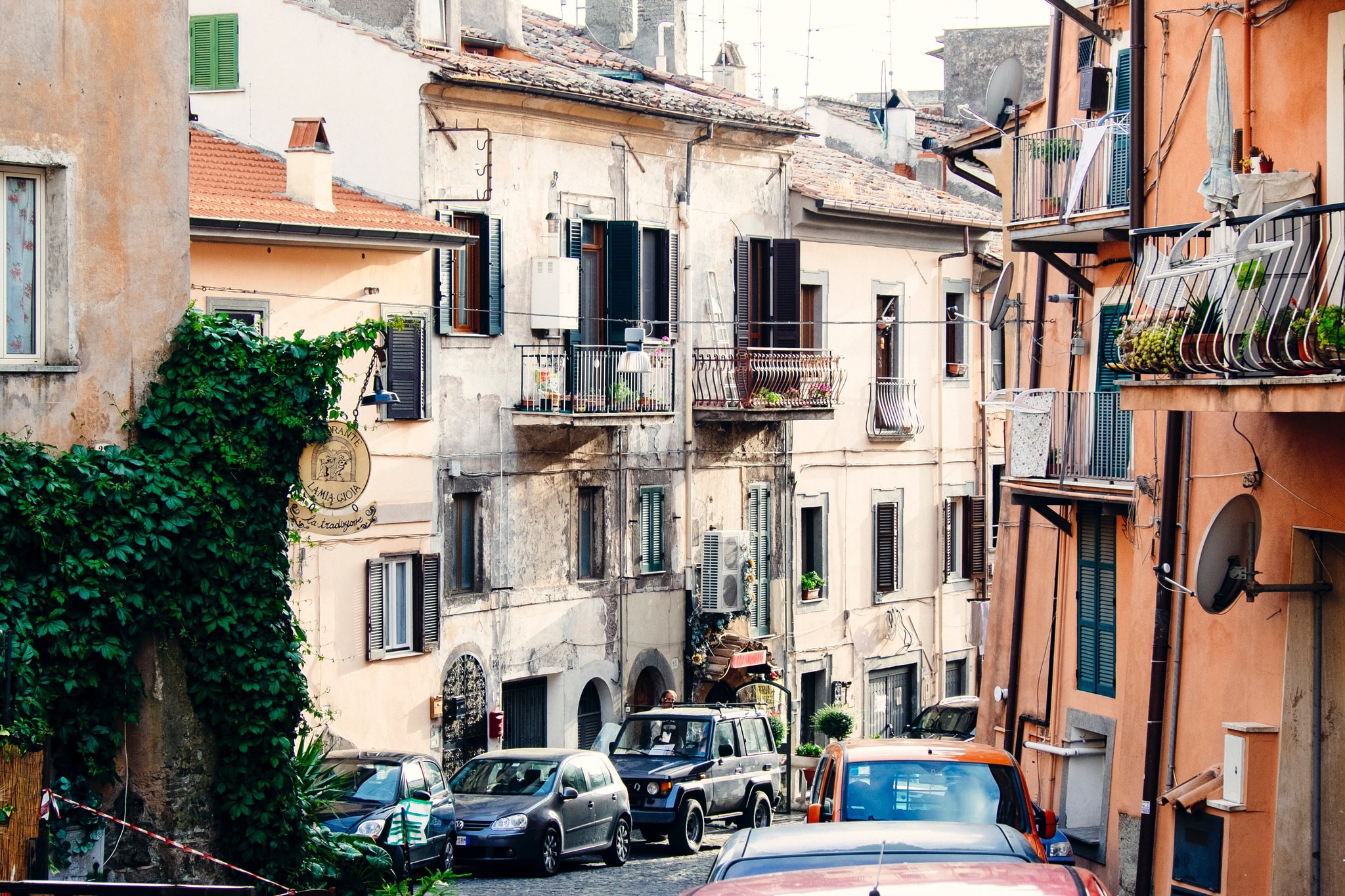}}
\subfigure[IMG-20161123-181711]{\includegraphics[scale=0.0484]{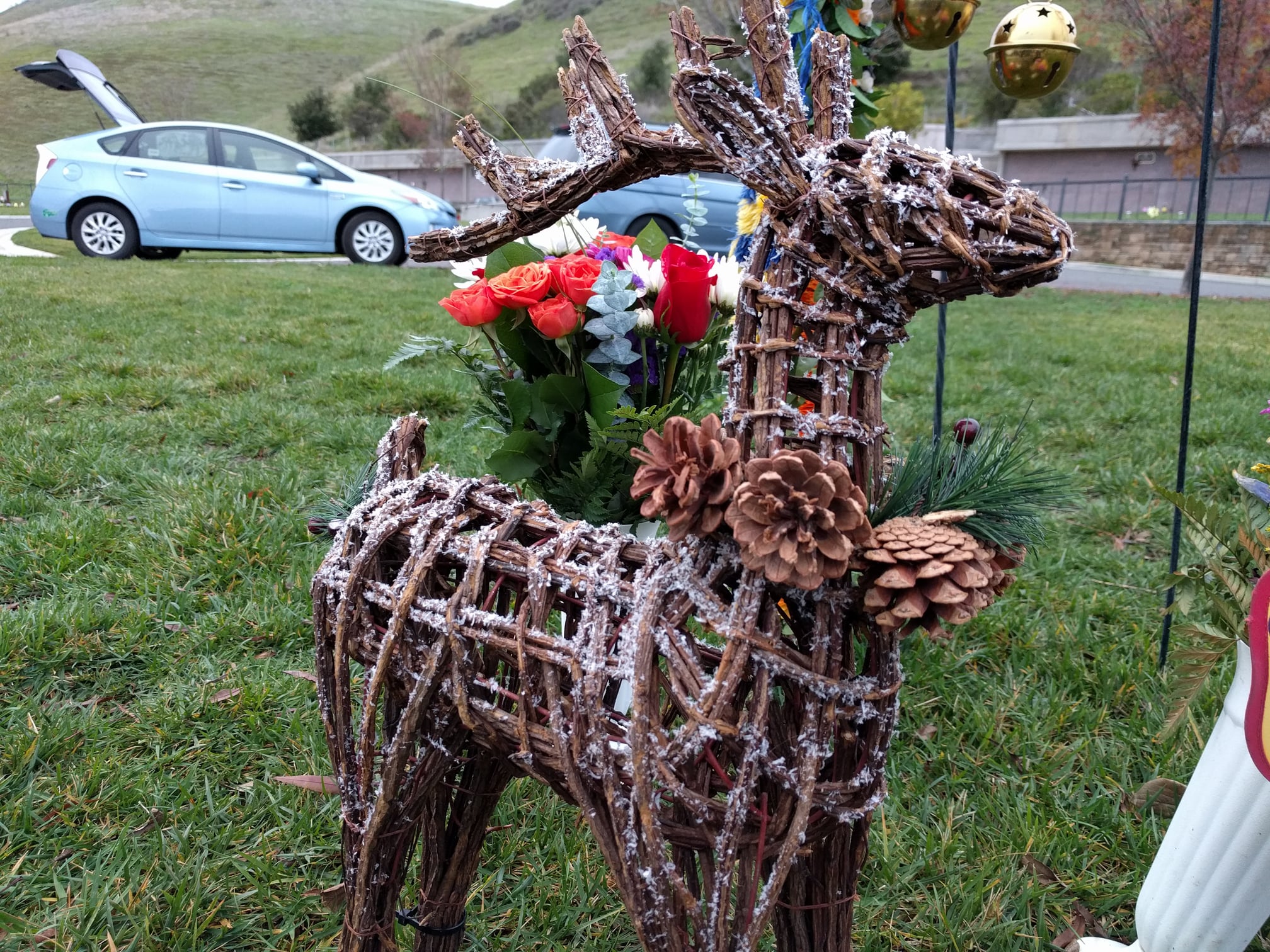}}
\subfigure[IMG-20170211-145346]{\includegraphics[scale=0.0484]{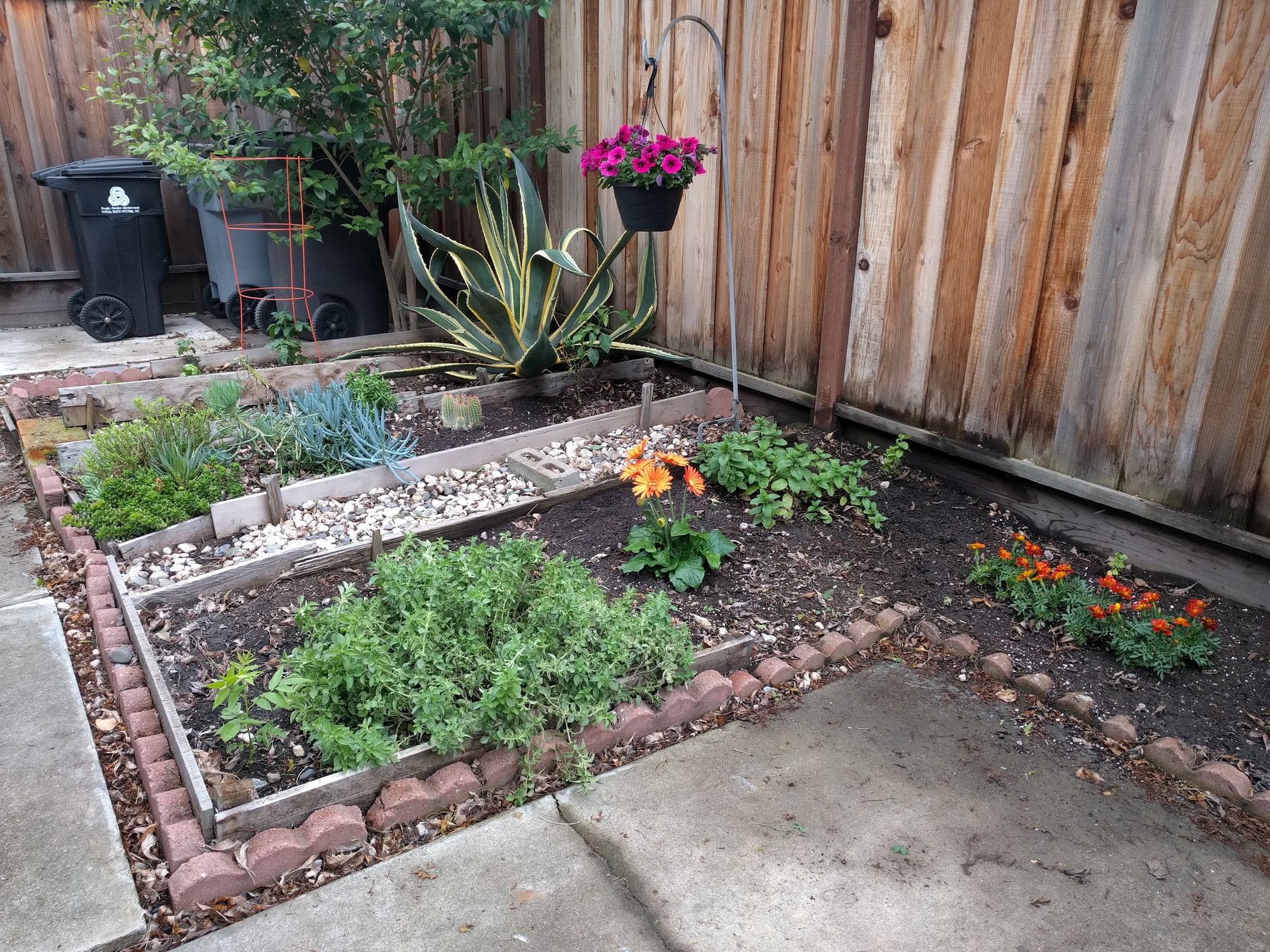}}
\caption{Four images sampled from CLIC test dataset}
\label{fig:images}
\end{figure}

\begin{figure}[t]
\centering
\subfigure[casey-fyfe-3340]{\includegraphics[scale=0.21]{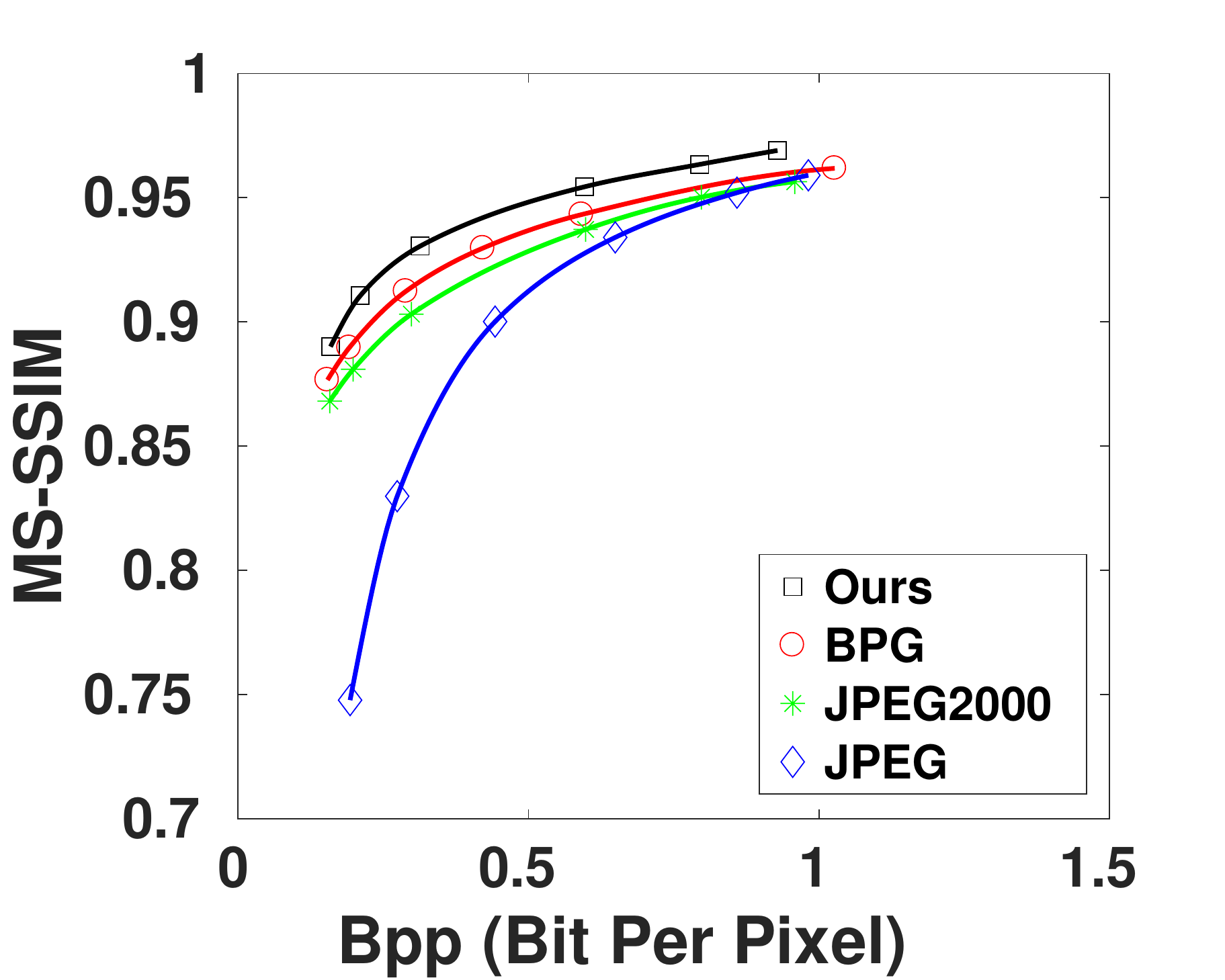}}
\subfigure[lou-levit-369]{\includegraphics[scale=0.2]{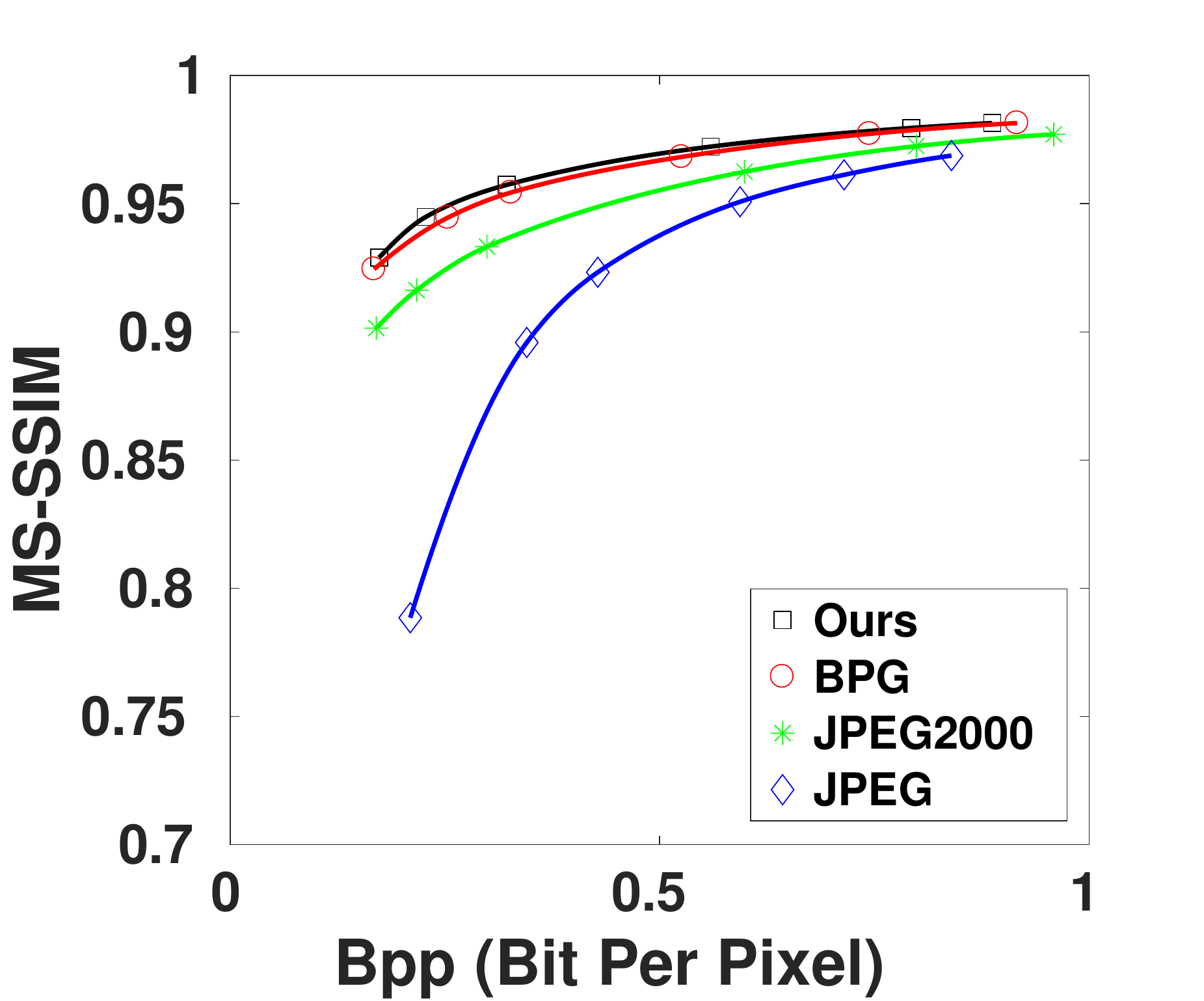}}
\subfigure[IMG-20161123-181711]{\includegraphics[scale=0.21]{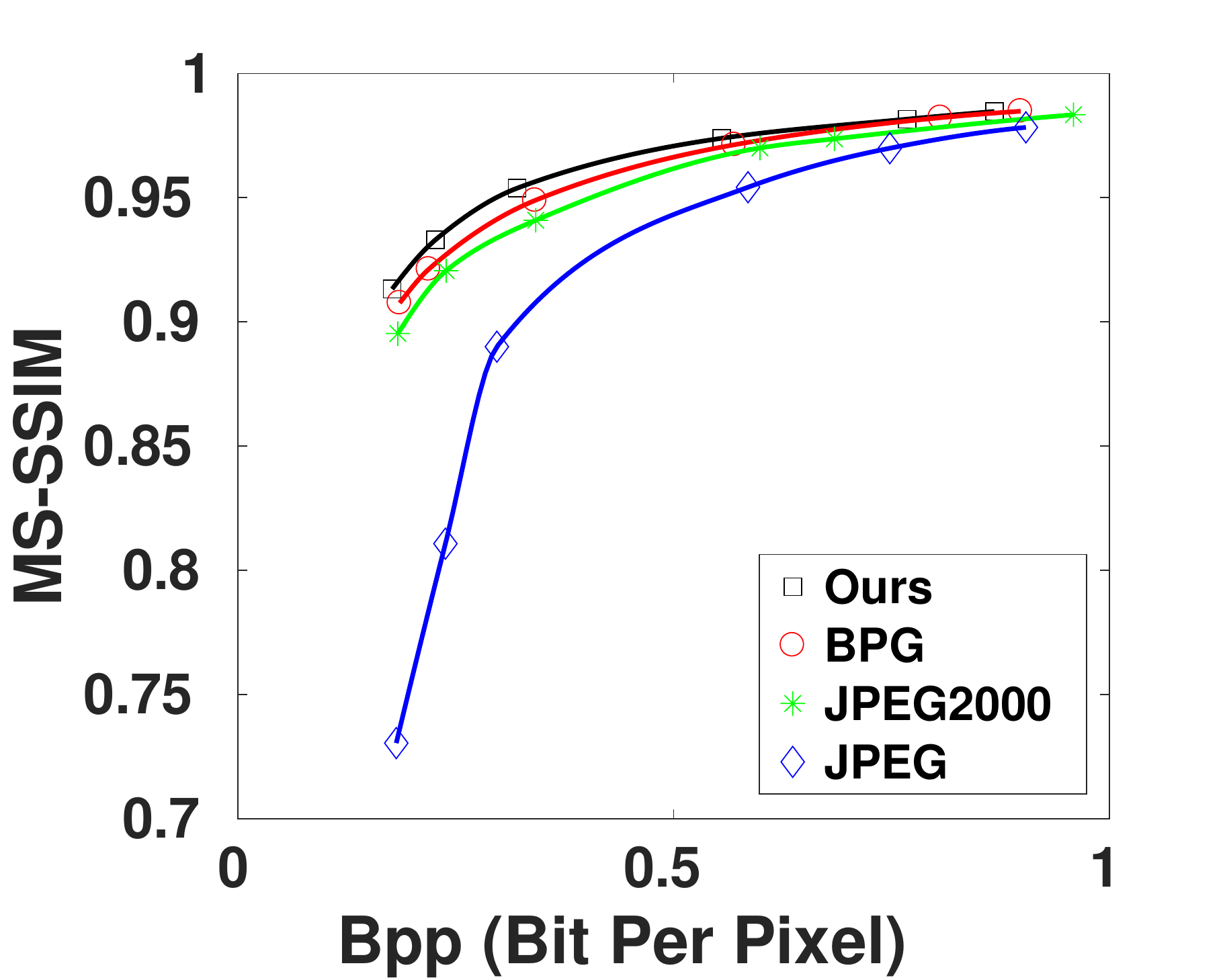}}
\subfigure[IMG-20170211-145346]{\includegraphics[scale=0.2]{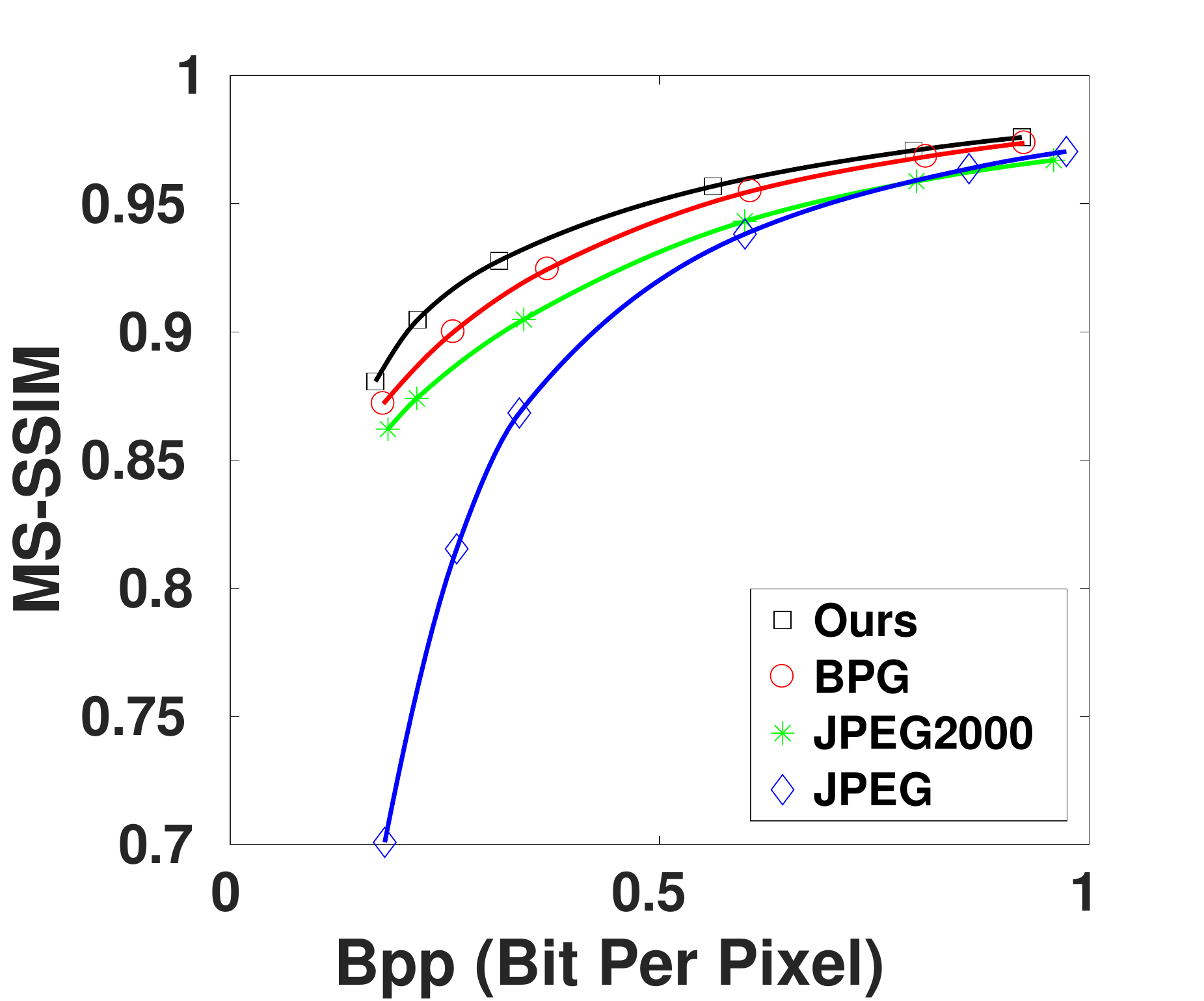}}
\caption{Compression performance on four images, all measured in RGB domain, compared with JPEG, JPEG2000 and BPG.}
\label{fig:curve}
\end{figure}

We evaluate our performance on the dataset released by CLIC and Kodak PhotoCD data set, and compare with existing codecs including JPEG, JPEG2000, and BPG.
Fig.~\ref{fig:Kodak} shows MS-SSIM performance over all 24 Kodak images and achieves averaged 7.81\% BD-Rate reduction over BPG\footnote{Given that BPG demonstrates the state-of-the-art coding efficiency, we mainly present the comparison against it.}.
Moreover, we have more impressive performance on CLIC test dataset. We select four typical images with different types from the dataset as test samples, as shown in Fig.~\ref{fig:images}. Finally, the BD-Rate has separately reduced by 33.54\%, 9.65\%, 13.31\% and 19.96\%, as illustrated in Fig.~\ref{fig:curve}
As can be seen from the results, our approach outperforms BPG and JPEG2000 for both overall performance and separate comparison using individual test image.

\section{Conclusion}
An end-to-end learning framework based deep image compression scheme is detailed in this work, with innovations among residual unit,
content adaptive fMaps, Lagrangian optimized rate-distortion adaptation, linear piecewise rate estimation, image visual quality enhancement with adversarial loss and perceptual loss included, and so on.  Our network coder is trained using the public data set released by CLIC2018.
Simulations are performed on independent images, resulting in impressive gains over the BPG and JPEG2000, both objectively and subjectively.
{\small
\bibliographystyle{ieee}
\bibliography{deepcoder}
}

\end{document}